# Cost Modeling and Projection for Stacked Nanowire Fabric

Naveen Kumar Macha, Mostafizur Rahman

Department of Computer Science & Electrical Engineering, University of Missouri Kansas City, MO, USA

E-mail: nmhw9@mail.umkc.edu, rahmanmo@umkc.edu

*Abstract*— To continue scaling beyond 2-D CMOS with 3-D integration, any new 3-D IC technology has to be comparable or better than 2-D CMOS in terms of scalability, enhanced functionality, density, power, performance, cost, and reliability. Transistor-level 3-D integration carries the most potential in this regard. Recently, we proposed a stacked horizontal nanowire based transistor-level 3-D integration approach, called SN3D [1][2] that solves scaling challenges and achieves tremendous benefits with respect to 2-D CMOS while keeping manageable thermal profile. In this paper, we present the cost analysis of SN3D and show comparison with 2-D CMOS, conventional TSV based 3-D (T3-D) and Monolithic 3-D (M3-D) integrations. In our cost model, we capture the implications of manufacturing, circuit density, interconnects, bonding and heat in determining die cost and evaluate how cost scales as transistor count increases. Since, SN3D is a new 3-D IC fabric, based on our proposed manufacturing pathway [2] we assumed complexity of fabrication steps as proportionality constants in our cost estimation model. Our analysis revealed SN3D have 86% and 74% reduction in area; 55% and 43% reduction in interconnect distribution and total interconnect length required; reduced metal layer requirement; and 70% and 68% reduction in total cost in comparison to 2-D CMOS and Monolithic 3-D (M3-D) integrations respectively.

## I. INTRODUCTION

Transistor-level 3-D integration is considered the most promising direction for replacing 2-D CMOS due to its density and performance benefits. Our proposal for transistor-level 3-D integration called stacked horizontal nanowire based 3-D IC fabric (SN3D), is based on stacked horizontal nanowires and uses architected features for circuit functionality, interconnection, and thermal management. Previously we reported huge gains with SN3D design [1-3]. In this paper, we focus on cost aspects and show a detailed comparison with 2-D CMOS, TSV based 3-D CMOS (T3-D) and monolithic 3-D (M3-D) integrations.

Fig. 1 shows core components of the fabric and overview of circuit mapped SN3D fabric. Stacked suspended horizontal nanowires (Fig. 1A) are the building blocks, which are prefabricated, and predopped. Architected fabric components: Gate-All-Around junctionless nanowire FETs (Fig. 1B), Common Contact (CC- Fig. 1C.i), Common Gate (CG- Fig. 1C.ii), Horizontal Bridges (HB- Fig. 1C.iii), Horizontal Insulation (HI-Fig. 1D) and Fabric Vias (FV- Fig. 1E) are formed onto these nanowires through material deposition techniques[4][5]. Active devices here are junctionless nanowire transistors that do not require any doping variation for Drain/Source/Channel regions. Previously we demonstrated junctionless device behavior experimentally [4][5]. The CG feature allows multiple junctionless devices to be gated with single input in vertical and/or horizontal directions. This enables optimal placement of common gated devices and minimizes inter-device connectivity requirements. The CCs provide common contact for stacked NWFETs, serve as an interconnection between adjacent transistors' source and drain contacts. Functionally, CG and CC carry common signals; however, physically they are composed of different material stacks according to channel-gate work-function and Ohmic Contact requirements respectively. The HI (Fig. 1D) provides isolation between adjacent Gate, Source and Drain contacts in the vertical direction. The HBs [4][5] serve dual purpose: connectivity and heat extraction. These Bridges along with CG and CC allow routability in 3-D and is very different from 2-D CMOS and other through silicon via based 3-D CMOS approaches where routing mostly take place in the 2-D plane and through vias that connect two layers/dies. Finally, fabric vias (Fig.1E) are used for input/output signals, and to carry excess routing to top metal layers, those maximize the routability in the SN3D fabric.

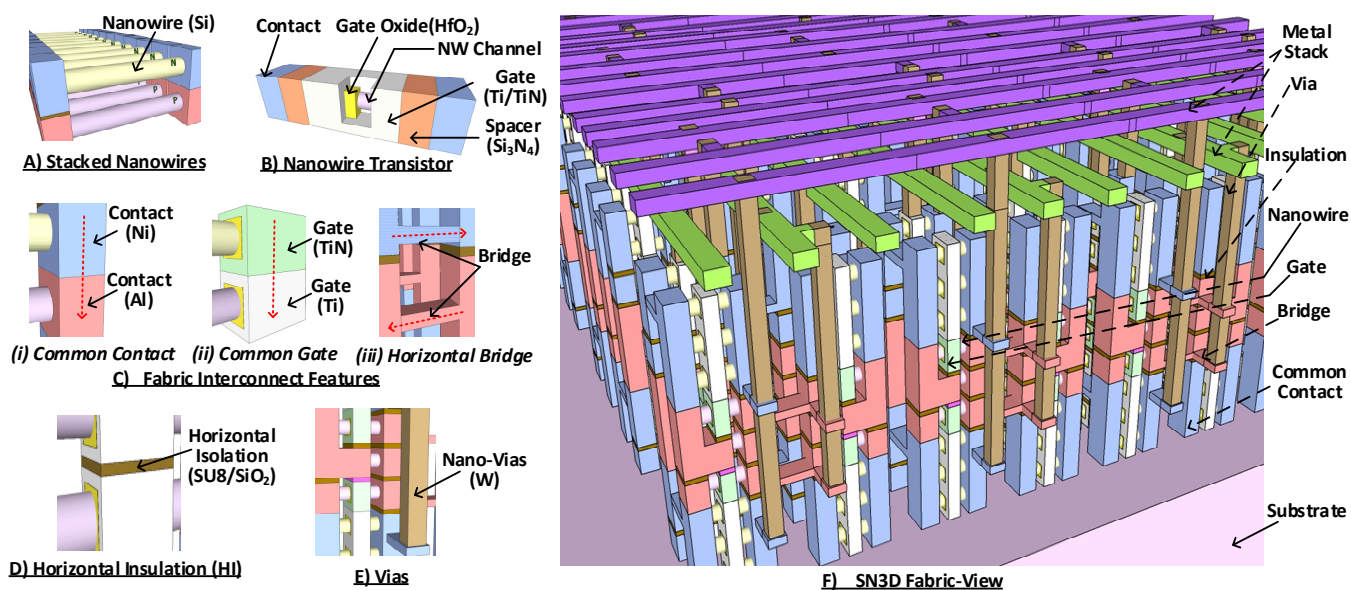

Fig. 1: (A) Stacked Nanowires; (B) Junction-less Gate All Around Nanowire Transistor; (C) Fabric-Interconnects, (i) Common Contact (CC), (ii) Common Gate (CG), (iii) Horizontal Bridge (HB); (D) Horizontal Insulation (HI); (E) Fabric-Vias; (F) SN3D Fabric-Overview

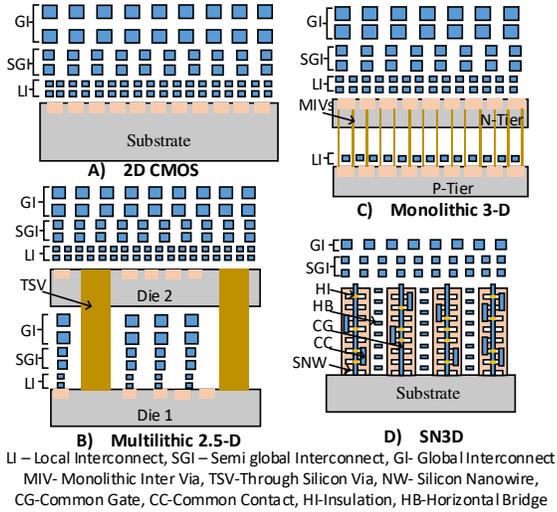

Fig. 2. Interconnect Hierarchy (A) 2D CMOS, (B) Through Silicon Via (TSVs) based 3-D IC, (C) Monolithic 3-D (M3D), (D) SN3D.

Fig.2 presents the conceptual view of the four integration approaches (2-D, T3-D, M3-D and SN3D) and illustrates their difference in terms of interconnection hierarchy. In 2-D CMOS, active devices are on the substrate at the bottom and all the routings are carried out on top metal layers (Fig.2 A). Similar are active device-layers and metal routing in conventional 3-D integration approaches (T3-D and M3-D), whereas, 3-D package and 3-D routing (Fig. 2 B & C) are through stacked prefabricated dies, and perforated thick inter-die vias (TSVs for T3-D and MIVs for M3-D) respectively. Here, the die stacking is limited (typically two layers) and relatively very few vertical interconnects replace the global interconnects. Whereas, SN3D is an integrated device-interconnect-fabric (fig. 2D and 1F), where, active devices are implemented in a stacked array of NWs, and 3-D interconnects are routed through architected fabric interconnect features. These fabric interconnect features replace the metal interconnects from local to global interconnect lengths (fig. 2D). Thus, SN3D offers a dense and fine-grained package of both devices and interconnects, yielding to huge footprint and interconnects length reduction. These imply circuits implemented through SN3D fabric offer increased functionality and improved performance at reduced power consumption[1-3]. However, any such benefits of an IC fabrication approach ultimately have to translate to cost benefits or at least to endurable cost trade-off, in order to adopt the approach to mainstream or research. This paper presents the benefits of SN3D implementation in cost form. The following section presents the cost model used to analyze the cost of SN3D implemented IC.

## II. COST MODEL

The cost model used for 2-D CMOS, T3-D, M3-D and SN3D is illustrated in the Fig. 3. As SN3D IC follows a new fabrication approach that does not have prior cost data, the early cost estimation model showed (Fig. 3.) is built upon the prior estimates of the major cost dictating aspects, such as, interconnect, die area, the complexity of process steps, and metal layers-requirement. Initially, the die area and the interconnect requirement is estimated based on SN3D design rules, guidelines and circuit design principles developed (Fig. 3.). Also, the 3D interconnects replacing the metal interconnects are expressed quantitatively. Later, interconnects and area estimated is input (Fig. 3.) to the metal layer estimation algorithm to calculate the number of metal layers required for the circuit system considered. Finally, the estimated die area and metal layer count serve as the independent variables in the cost function developed, where, the fabrication complexity i.e., the total number of different process steps required, are parameterized as proportionality parameters. Also, to note, the cost model presented is applicable to four of the fabrication approaches considered, because the underlying fabrication processes are same for all, therefore the model enables the fair relative comparison of SN3D cost estimates with the rest. Towards which, the next sections present the scheme for interconnect, die area and metal layers estimation.

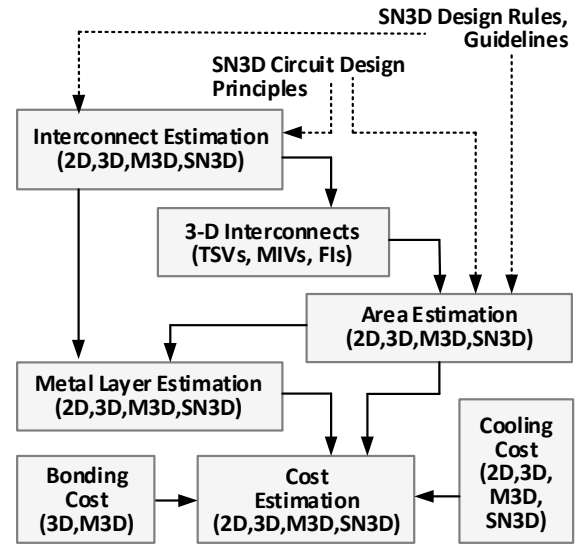

Fig. 3 Cost Estimation Model

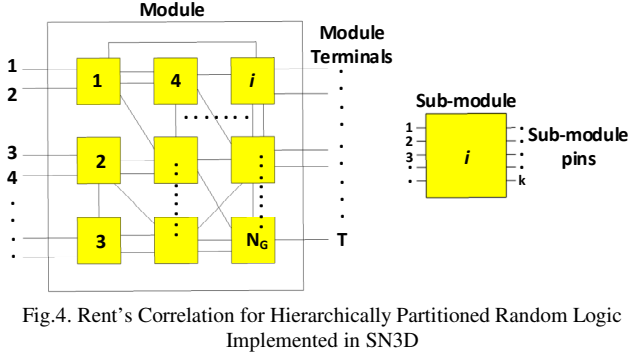

Fig.4. Rent's Correlation for Hierarchically Partitioned Random Logic Implemented in SN3D

## A. Interconnect Estimation for SN3D

Interconnect projection for a placement of logic gates is an important variable for early estimation of wiring space requirement, metal layer requirement, delays, power dissipation, and cost. For that, this section shows the hierarchical partitioning and placement strategy considered for SN3D fabric and estimates the interconnect density for circuits implement in SN3D. The interconnect estimates obtained would be used in later sections for estimation of metal layers, and cost (Fig.3). For a hierarchically partitioned circuit system (as shown in Fig.4) implemented in SN3D, the number of metal interconnect terminals ($T_m$) required by a module containing $N_G$ blocks is given by Rent's [6] empirical expression as (it is applicable at all levels of hierarchy)

$$T_m = k N_G^p \quad (1)$$

Where $k$ is the average number of terminals (input and output) per each block (Fig.4), and $p$ is the rent's constant which specifies the complexity of the design considered. From this, the total number of interconnects ($I_T$) required for a circuit system is proportional to the number of terminals and is given by the expression [7]

$$I_T = \alpha k N_G (1 - N_G^p) \quad (2)$$

Where $\alpha$ is the fraction of the on-chip terminals that are sink terminals, and it is related to average fanout $f.o.$, as $\alpha = f.o./f.o.+1$. In SN3D approach, a significant number of these interconnects are replaced by fabric interconnects, which reduces the interconnect routing on top metal layers. These interconnects reduction is expressed quantitatively next; a modified rent's expression [9] and an interconnect density function [8] is formulated for SN3D. Consider a circuit design with $N_G$ gates as represented by Fig.4, they can be equally distributed in to $n$ nanowire transistor layers (we consider $n$ =10 in this paper) in the SN3D fabric. This partitioning scheme is represented in Fig.5A, here each layer incorporates $\frac{N_G}{n}$ gates. The number of terminals emanating from each layer ($T_i$) due to such partition can be given by Rent's expression as

$$T_i = k \left(\frac{N_G}{n}\right)^p \quad (3)$$

Where $i$ is the layer's number and $n$ is the total number of layers. Summation of $T_i$ over all the layers (Fig.5A) gives the total number of terminals $T_{SN3D}$ available in SN3D over all the layers.

$$T_{SN3D} = \left(\sum_{i=1}^{n} T_i\right) = nk \left(\frac{N_G}{n}\right)^p \quad (4)$$

Next, the Fig.5B represents partitioning and placement scheme in SN3D fabric, where, the total terminals ($T_{SN3D}$) emanating are utilized in two distinct ways: $T_{feat}$ terminals for 3-D interconnect features specific to SN3D, such as CG, CC and HB (Fig.1 and 2D); and $T_m$ terminals for metal layer interconnects. Hence, $T_{SN3D}$ is comprised of (Fig.5B)

$$T_{SN3D} = T_m + T_{feat} \quad (5)$$

Now, from $eq(1), eq(4)$ and $eq(5)$

$$T_{feat} = nk \left(\frac{N_G}{n}\right)^p - k N_G^p$$

$$T_{feat} = n(1 - n^{p-1}) k \left(\frac{N_G}{n}\right)^p$$

From which, the average number of terminals consumed by fabric's 3-D interconnects features per each nanowire layer is

$$T_{feat,i} = \frac{T_{feat}}{n} = (1 - n^{p-1}) k \left(\frac{N_G}{n}\right)^p$$

Similarly, the number of terminals contributing top metal layer interconnects per each nanowire layer ($T_{m,i}$) is the difference of total number of terminals per each layer $T_i$ ($eq(3)$) and layer's fabric interconnect terminals ($T_{feat,i}$)

$$T_{m,i} = T_i - T_{feat,i}$$

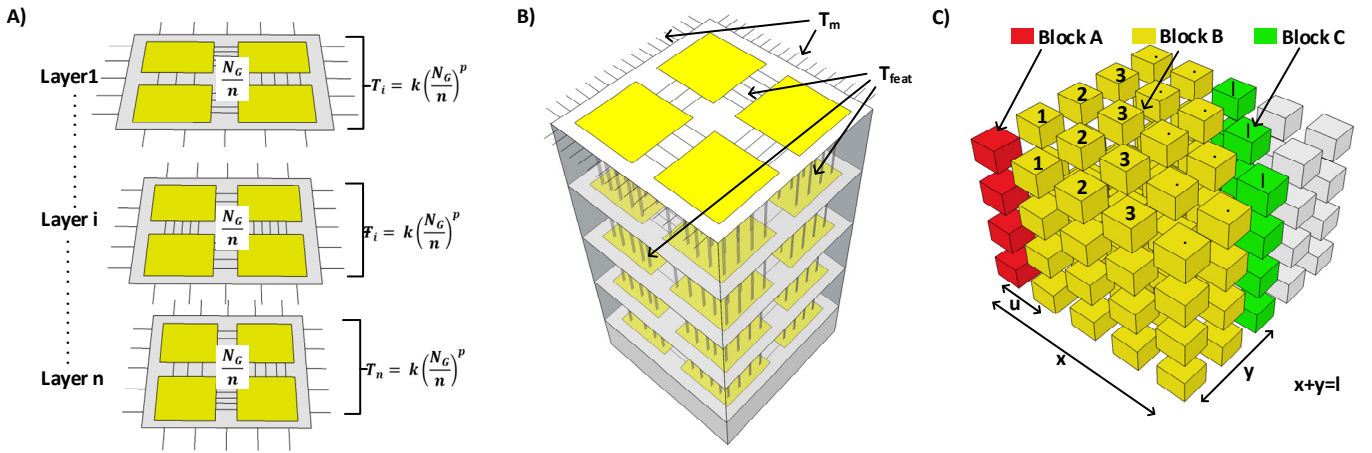

Fig.5. (A) Partitioning scheme for SN3D where $N_G$ gates are equally distributed into n layers; (B) Partition and Placement scheme in SN3D fabric leading to two distinctly utilized interconnect terminals ($T_m$ and $T_{feat}$) for the logic blocks; (C) Regular square arrayed placement of logic blocks in SN3D fabric for deriving interconnect distribution

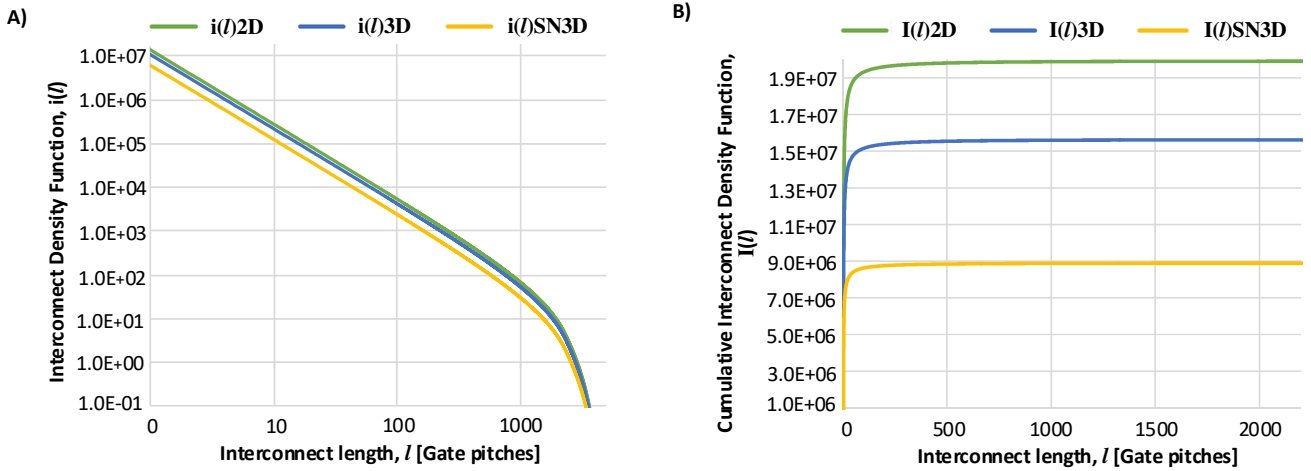

Fig.6. Fig.6. Interconnect Distributions calculated for 2-D, 3-D, and SN3D integration approaches: A) Interconnect Density Function $i(l)$; B) Cumulative Interconnect Density Function $I(l)$.

$$T_{m,i} = k\left(\frac{N_G}{n}\right)^p - (1-n^{p-1})k\left(\frac{N_G}{n}\right)^p$$

$$T_{m,i} = kn^{p-1}\left(\frac{N_G}{n}\right)^p$$

Comparing $T_{feat,i}$ and $T_{m,i}$ with its Rent's equivalent for each layer ($eq$ (3)), the effective average number of terminals per each block consumed for fabric interconnects ($K_{SN3D,feat}$) and effective average metal interconnects ($K_{SN3D,m}$) are

$$K_{SN3D,feat} = k(1-n^{p-1}) \text{ and}$$
$$K_{SN3D,m} = k(n^{p-1}) \quad (6)$$

Now, the total number of fabric interconnects can be expressed quantitatively by replacing $k$ with $K_{SN3D,feat}$ in $eq$(2)

$$I_{T,feat} = ak(1-n^{p-1})N_G(1-N_G^p) \quad (7)$$

Similarly, the metal interconnects of different lengths can be calculated by replacing $k$ with $K_{SN3D,m}$ in an interconnect estimation model. Fig.5C shows a regular square arrayed placement of a hierarchically partitioned system in SN3D, where each cube represents a logic block, and the spacing between the adjacent blocks is the unit distance called gate pitch (u). All interconnects between logic blocks are assumed at their manhattan grid distances [7][8] in the square array. As an example, elements numbered 2 in block B (Fig.5C) are at distance of 2 units (2u) from block A, similarly, elements in block C are at distance of $l$ (i.e., x+y) from elements in block A. For such a placement, ref. [8], gives a stochastic wire length distribution function $i(l)$; $i(l)$ is a continuous interconnect density function which gives number of interconnects available at a given length $l$. The closed form analytical expression $i(l)$ is a function of $N_G$, $l$, $k$ and $p$ variables,

$$i(l) = f(N_G, l, k, p)$$

These variables capture the characteristics of the circuit system considered and the silicon floor architecture. $N_G$ considers size of the logic design i.e., number of logic blocks in the circuit system; $p$ is the rent's coefficient which defines the complexity of the circuit considered; $l$ is the manhattan distance in multiples of u (u defines architecture of the silicon floor i.e., it is different for 2-D, T3-D, M3-D and SN3D); and $k$ is the average number of terminals per each block in the circuit system. As discussed in $eq$ (5), the reduction in metal interconnects due to fabric interconnects is expressed through reduction in $K_{SN3D,m}$. Thus, employing $K_{SN3D,m}$ for k [9], the modified expression for $i(l)$ [8] in SN3D is,

Region I:
$1 \le l \le \sqrt{N_G}$

$$i_{SN3D}(l) = \frac{\alpha(kn^{p-1})}{2}\tau\left(\frac{l^3}{3} - 2\sqrt{N_G}l^2 + 2N_G l\right)l^{2p-4}$$

Region II: $\sqrt{N_G} \le l \le 2\sqrt{N_G}$

$$i_{SN3D}(l) = \frac{\alpha(kn^{p-1})}{6}\tau\left(2\sqrt{N_G} - l\right)^3 l^{2p-4} \quad (8)$$

Where $\alpha$ is the fraction of the on-chip terminals that are sink terminals, and is related to average fanout $f.o.$, as

$$\alpha = \frac{f.o.}{f.o+1}$$

The normalization factor $\tau$ is given as [8]

$$\tau = \frac{2N_G(1-N_G^{p-1})}{\left(-N_G^p\frac{1+2p-2^{2p-1}}{p(2p-1)(p-1)(2p-3)} - \frac{1}{6p} + \frac{2\sqrt{N_G}}{2p-1} - \frac{N_G}{p-1}\right)}$$

Now, integration of $i(l)$ over a range of interconnect lengths gives the total number of interconnects available over that range, which is named as cumulative interconnect density function $I_{SN3D}(l)$ [8],

$$I_{SN3D}(l) = \int_1^l i_{SN3D}(\varsigma)\,d\varsigma \quad (9)$$

Fig. 6 shows the interconnect distribution $i(l)$ and cumulative Interconnects distribution $I(l)$ functions evaluated for SN3D, M3-D and 2-D designs. It can be noticed that SN3D shows huge reduction in interconnect distributions (Fig.6) from local to global interconnects range due to the availability of fine-grained 3-D fabric routing features (Fig.1 & 2D) from device to system level. This impact is pictorially depicted in Fig.7: the quantity of 3-D interconnects is given by $I_{T,feat}$ $eq$ (7) which is represented in the figure by interconnect lines among the blocks; and the reduced metal layer interconnects is given by $I_{SN3D}(2\sqrt{N_G})$ $eq$ (9) which is represented as metal lines on the top. The reduction in total interconnects $I_{SN3D}(2\sqrt{N_G})$ over $I_{2D}(2\sqrt{N_G})$ is 54% (this can be further enhanced by exploring the efficient routing options in SN3D fabric). This reduction reduces the average interconnect length and total wiring requirement of the chip, consequently alleviates the metal layers requirement, wire delays, power loss and cost.

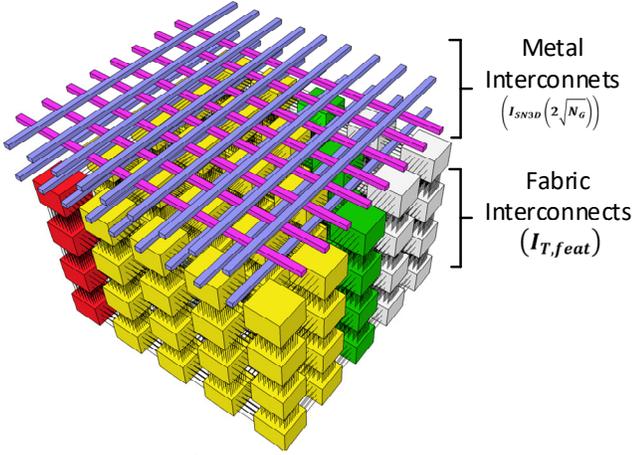

Fig. 7. Potentials of SN3D fabric: Dense placement strategy for devices/logic blocks; Increase in interconnect resources due to Fabric's 3-D interconnects ($I_{T,feat}$); Reduction in metal interconnect requirement

### B. Die Area Estimation

Die area for gate-limited designs is proportional to the total number of gates available in the design [10], accordingly, it can be formulated for 2-D, T3-D, M3-D as

$$A_{2D} = N_G A_{G,2D}$$
$$A_{3D} = N_G A_{G,3D} + N_{TSV} A_{TSV}$$
$$A_{M3D} = N_G A_{G,M3D} + N_{MIV} A_{MIV}$$

Where $N_G$ is the total number of gates in the design and $A_G$ is the average area of each gate. The average gate area in 2-D CMOS ($A_{G,2D}$) is taken from [10] as 3125 $\lambda^2$, correspondingly, assuming an efficient 2 layer 3-D integration, the average gate area for T3-D and M3-D is considered half the 2-D CMOS gate area (i.e., $A_{G,3D} = A_{G,M3D} = \frac{A_{G,2D}}{2}$). Additionally, to note from above expressions, T3-D and M3-D suffer from area overhead due to TSVs and MIVs respectively. Here, $N_{TSV}$ and $N_{MIV}$ are the number of TSVs and MIVs required respectively; $A_{TSV}$ and $A_{MIV}$ are the block out area for each TSV and MIV respectively. $N_{TSV}$ is estimated from [10], and $N_{MIV}$ for a transistor-level-monolithic 3-D design could be expressed from $eq$ (6) and $eq$ (2) as (M3-D is limited to 2 tier design)

$$I_{T,MIVs} = ak(1 - 2^{p-1})N_G(1 - N_G^p)$$

Similarly, die area for SN3D is given by

$$A_{D,SN3D} = N_G A_{G,SN3D}$$

Based on our previous SN3D layout designs [1-3], the average gate area $A_{G,SN3D}$ calculated for SN3D is 432$\lambda^2$. Fig.8 represents single gate layout [2] in SN3D consuming the footprint of only one NW transistor (other transistors are implemented in subsequent nanowires on the bottom layers) plus the footprint for fabric vias, and HBs. It is to be noticed that block out area for fabric vias is incorporated in the gate area of SN3D.

Thus, solving the given die area expressions, Fig.9 shows the die area estimates for 2-D CMOS, 3-D, M3-D and SN3D approaches for 5,10 and 20 million logic gate designs. From the figure, SN3D shows a huge reduction in area compared to other approaches; the reduction is 86% and 74% with respect to 2-D CMOS and M-3D respectively. Subsequent sections will use these area estimates for metal layer estimation and cost projection.

### C. Metal layer Estimation

The metal layer estimation algorithm [11] [10] is based on the iterative bottom-up placement of interconnect distribution onto the available routing area in successive metal layers. The total interconnect routing length available on a metal layer is given by [11]

$$L_{av,i} = \frac{\mu_i A_{D,SN3D} - A_{vias,i}}{w_i}$$

Where $A_{D,SN3D}$ is the SN3D die area, $i$ indicates the metal layer count, $\mu_i$ is layer's routing efficiency, $A_{vias,i}$ is the layer's total via block out area and $w_i$ is the layer's metal wire pitch. $A_{vias,i}$ is estimated as [11]

$$A_{vias,i} = 2A_{v,i}(N_G f.o - I(l_i))$$

Where $A_{v,i}$ is block out area of single via on layer $i$, $N_G$ is the total gate count, $f.o$ is the average fanout per each gate and $I(l_i)$ is the cumulative interconnect density which gives the total number of interconnects routed until current layer. Here, $(N_G f.o)$ is the total number of fanout interconnects available in the design, thus, $(N_G f.o - I(l_i))$ gives total number of fanout interconnects routed above current metal layer $i$ [11], twice this number gives the approximate number of vias passing through current metal layer [11]. Fig.9 illustrates the metal layer estimation algorithm. Starting from bottom metal layer, the interconnects available through interconnect distribution function are routed iteratively until the layer's available routing length $L_{av,i}$ is deplete. The subsequent interconnects are routed similarly on the next metal layers as depicted in the Fig.9. The upper bound condition for iterative routing of interconnects in a metal layer is

$$\chi L(l_i) - \chi L(l_{i-1}) \leq L_{av,i}$$

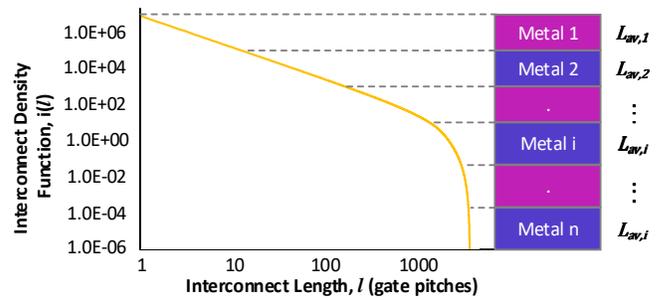

Fig.9. Metal Layer Estimation Algorithm

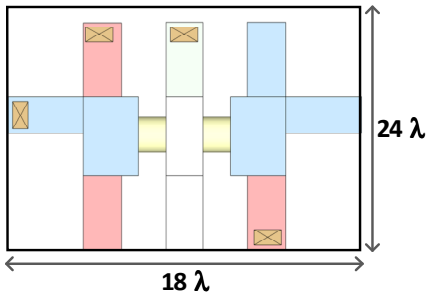

Fig.8. Layout of SN3D Standard Cell

Where $L(l_i)$ is the cumulative interconnect wire length until the $i^{th}$ layer, and $\chi = \frac{4}{f.o+3}$ is the factor that accounts for fraction of interconnects shared with in a common net [11]. Upon completion of all the interconnects available in interconnection distribution, the routing algorithm terminates, and the value of $i$ gives the estimation of metal layers required for the circuit system considered. Table.1 presents the results of metal layer estimation for 2D, T3-D, M3-D and SN3D approaches for 5, 10 and 20 million logic gate designs. SN3D shows 2.5x reduction in metal layer requirement, which would reduce the cost of metal layers, that is discussed in the next section.

TABLE. 1
METAL LAYER ESTIMATION

| Ng | 2-D CMOS | TSV 3D | M 3D | SN3D |
|---|---|---|---|---|
| 5 M | 5 | 5 | 3 | 3 |
| 10 M | 6 | 5 | 4 | 3 |
| 20 M | 7 | 6 | 5 | 4 |

*D. Cost Approximation*

Cost of a chip is comprised of its constituent costs: die cost, metal layers cost, cooling cost and bonding cost (package costs is not considered).
$$C = C_{die} + C_{metal} + C_{cooling} + C_{bonding} \quad (11)$$
Projection of this constituent costs for a fabrication approach needs both prior cost data from foundry and the design phase information available. As SN3D is a new fabrication approach, which has no previous cost data from the foundry, we have developed a cost model which best captures the design phase information available and presents a forecast proportional to actual cost, those, it enables faithful comparison of SN3D cost with 2-D, T3-D and M3-D approaches. The design phase information like die area ($A_{2D}$), metal layers ($n_m$) and fabrication process steps requirement, serve as the determinants. Accordingly, the general expression for die cost and metal cost can be expressed as
$$C_{die} + C_{metal} = (c_{pd}A_{Die}) + (c_{pm}n_m A_{Die}) \quad (12)$$
Where $A_{Die}$ and $n_m$ (estimated is previous sections) are the variables dependent on physical design of the circuit system, $c_{pd}$ and $c_{pm}$ are the proportionality constants parameterized corresponding to the process steps involved in fabrication. $c_{pd}$ is a parameter proportional to process steps involved in the fabrication of active devices on die and $c_{pm}$ is the parameter proportional to process steps involved in fabrication of metal layers on top of the die. Next, we formulate these parameters for different fabrication approaches.

The sequential processes involved in any IC fabrication can be classified into five major process steps, photolithography, diffusion, deposition, etching, and implantation. Next, we quantize and parameterize these process steps in terms of an arbitrary cost. Let $k_c$ be an arbitrary cost consumed by unit area of silicon chip when subjected to each of the five processes. This is conveyed in the Fig.10A, where, a unit area of substrate is subjected to one photolithography, one diffusion, one implantation, one deposition and one etching process steps. Thus, $k_c$ can be expressed into constituent cost constants of

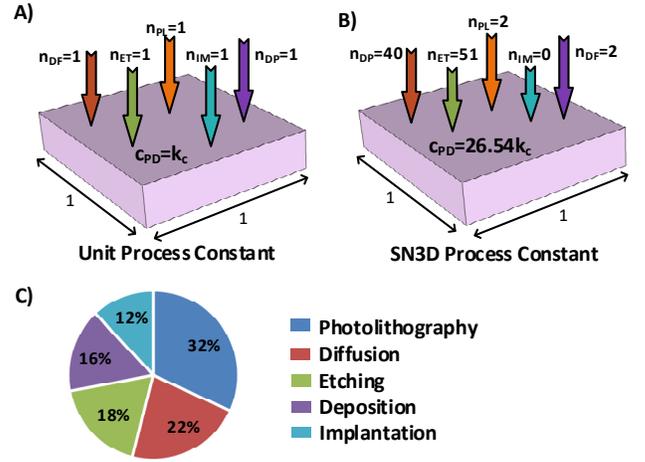

Fig. 10. Parameterizing Process Steps A) Unit process parameter for cost, B) SN3D process parameter for cost, C) Relative Cost of the process steps

photolithography $k_{PL}$, diffusion $k_{DF}$, deposition $k_{DP}$, etching $k_{ET}$ and implantation $k_{IM}$.
$$k_c = k_{PL} + k_{DF} + k_{ET} + k_{DP} + k_{IM}$$
Now, Fig.10C [12] depicts the relative cost of these major process steps involved in the semiconductor industry. From these relative cost statistics (Fig.10C) [12], the above constituent cost constants can be appropriately deduced to fraction of $k_c$ i.e., $k_{PL} = 0.32k_c; k_{DF} = 0.22k_c; k_{ET} = 0.18k_c; k_{DP} = 0.16k_c$; and $k_{IM} = 0.12k_c$

Next, determining the number of different process steps from the process sequence [2] [13-17] of a particular fabrication approach would give cost per unit area ($c_p$) for that approach.
$$c_p = n_{PL}.k_{PL} + n_{DF}.k_{DF} + n_{ET}.k_{ET} + n_{DP}.k_{DP} + n_{IM}.k_{IM}$$
$$c_p = (0.32n_{PL} + 0.22n_{DF} + 0.18n_{ET} + 0.16n_{DP} + 0.12n_{IM})k_c \quad (13)$$
Where $n_{PL}$, $n_{DF}$, $n_{ET}$, $n_{DP}$ and $n_{IM}$ are the number of photolithography, diffusion, etching, deposition, and implantations steps required respectively. For example, Fig.10B conveys the number of different process steps required for SN3D die (Table.2) [2] [13], therefore, $c_p$ for SN3D die ($c_{pd}$) is calculated from $eq(13)$ as $26.54k_c$. Similarly, Table.2 presents the number of different process steps required per unit area of 2D CMOS [14], T3-D [15], M3-D [16] and SN3D [2] [13] dies, and for unit area of a metal layers [14]. It is to be noticed that T3-D and M3-D require more than twice the 2-D process steps because two stacked dies would incur equal but separate process steps and the process steps excess to twice the 2-D CMOS are to drill TSVs and MIVs through the dies. Substituting this process steps count in $eq(13)$, and from $eq(12)$ and (11), the final cost expressions formulated for 2-D, T3-D, M3-D and SN3D are
$$C_{2D} = 6.26k_c A_{2D} + 2k_c n_m A_{2D} + C_{cooling}$$

TABLE 2
PROCESS STEPS

| Process | 2D [14] | 3D/M3D [15-17] | SN3D [2][13] | Metal [14] |
|---|---|---|---|---|
| Photolithography | 9 | 19 | 2 | 2 |
| Diffusion | 4 | 8 | 2 | - |
| Implantation | 7 | 14 | - | - |
| Deposition | 4 | 10 | 40 | 4 |
| Etching | 5 | 13 | 51 | 4 |

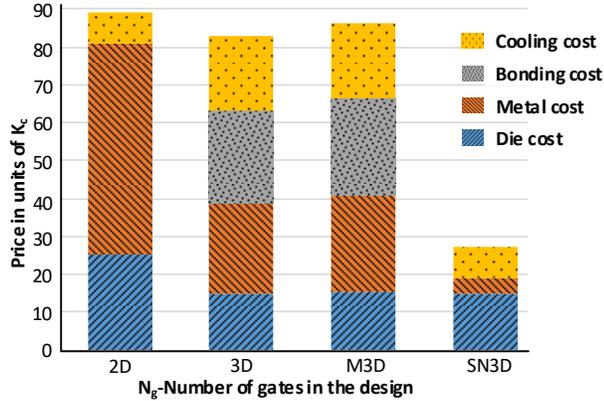

$$C_{T3D} = 7.26k_c A_{3D} + 2k_c n_m A_{3D} + C_{bonding} + C_{cooling}$$
$$C_{M3D} = 7.26k_c A_{M3D} + 2k_c n_m A_{M3D} + C_{bonding} + C_{cooling}$$
$$C_{SN3D} = 26.54k_c A_{SN3D} + 2k_c n_m A_{SN3D} + C_{cooling}$$

Where 6.26, 7.26, 7.26, and 26.54 are the die process parameters ($c_{pd}$) calculated for 2-D, T3-D, M3-D and SN3D respectively; whereas, metal layers' process parameter $c_{pm}$ is same (i.e., 2) for all the approaches because metal layer process steps are identical in all the approaches. $A_{2D}, A_{3D}, A_{M3D},$ and $A_{SN3D}$ are the die areas estimated in section B for 2-D, T3-D, M3-D and SN3D respectively, and metal layer estimates ($n_m$) for all the approaches are from Table.1. Hence, by evaluating the above expressions, SN3D cost is estimated and compared to conventional fabrication approaches.

Fig.11 shows the cost estimates evaluated for four of the integration approaches along with its components costs. SN3D cost estimate shows huge saving due to following reasons. First, though the number of process steps increases in SN3D due to voluminous package of devices (this is reflected in process constant 26.54), huge reduction in the die area reduces the die cost. Moreover, as noted in SN3D process count in Table.2 and process cost-statistics from Fig.10C, SN3D shifts the processes to relatively cheaper deposition and etching steps compared to extreme lithography steps needed in sub nanometric regime in conventional approaches. Second, reduction in metal interconnects and metal layer requirement alleviates the metal layer cost for SN3D. Third, SN3D does not require any bonding cost (Fig.11), whereas 3-D and M3-D incur significant bonding cost. The bonding costs presented are estimated based on relative cost data from [10]. Fourth, the cooling cost is linearly proportional to the temperature of the chip [10]; 3-D and M3D require relatively high cooling cost because of the extreme working temperatures due to minimum heat escape path in the stacked dies. Whereas SN3D owing to its fine-grained thermal management scheme heats up less [18] compared to other three, therefore, it incurs a low cooling cost.

## III. CONCLUSION

We have presented in detail an approach for relative cost estimation of different fabrication methodologies which is based on the quantitative estimation of major cost dictating aspects, and compared the cost of SN3D IC with 2-D CMOS, T3-D and M3-D (Fig.12) costs. This cost evaluation methodology is generic and it is applicable for relative cost estimation of any fabrication methodologies, whereas the parameterized constants change according to the process complexity (number of different process steps) and can be derived from $eq$ (13). Such kind of cost estimation would enable a fair comparison of cost estimates at design phase, and, as it is in units of an arbitrary cost ($k_c$) which serves as a base enabling consistent comparison between different fabrication approaches, the cost estimated would be proportional to the actual cost. Our results show SN3D consumes 70%, 67% and 68% lower cost than 2-D, T3-D and M3-D respectively, proving it to be promising 3-D integration direction for research.